\documentclass[a4paper,11pt]{article}
\usepackage{pos}

\def\bea{\begin{eqnarray}}
\def\eea{\end{eqnarray}}
\def\nn{\nonumber}

\def\lmatrix{\left(\begin{array}}
\def\rmatrix{\end{array}\right)}
\def\gsim{\mathrel{\rlap{\lower4pt\hbox{\hskip1pt$\sim$}}\raise1pt\hbox{$>$}}}
\def\lsim{\mathrel{\rlap{\lower4pt\hbox{\hskip1pt$\sim$}}\raise1pt\hbox{$<$}}}
\def\bi{\begin{itemize}}
\def\ei{\end{itemize}}
\def\msbar{\overline{\rm MS\kern-0.5pt}\kern0.5pt}
\def\rho{\varrho}

\title{Spin-1 fields and RG flows in 4 dimensions}

\author{Daniel Nogradi}

\affiliation{Eotvos University, Department of Theoretical Physics\\
Pazmany Peter setany 1/a, Budapest 1117, Hungary}

\emailAdd{nogradi@bodri.elte.hu}

\abstract{The most general local, classically scale invariant, perturbatively renormalizable, 
globally $SU(N)$ invariant Lagrangian
is constructed for spin-1 fields in 4 dimensions. The total number of independent couplings is 7 and 
the 1-loop $\beta$-functions are computed in the $\msbar$ scheme. A number of asymptotically free
RG flows are identified corresponding to non-trivial QFTs. None of these are gauge theories. The
details of the large-$N$ limit are also worked out and it is shown that the RG phase space is
qualitatively similar for all $N>5$ including the $N\to\infty$ limit. 
}

\FullConference{
 The 38th International Symposium on Lattice Field Theory, LATTICE2021
  26th-30th July, 2021
  Zoom/Gather@Massachusetts Institute of Technology
}

\begin{document}
\maketitle

\section{Introduction}
\label{introduction}

In this work a straightforward QFT question is asked: what type of QFT can describe interacting,
asymptotically free spin-1 (vector) fields in 4 dimensions? If gauge invariance is imposed
Yang-Mills theory is unique and well-known, hence we do not require gauge invariance here only a {\em global} $SU(N)$
invariance, beyond locality, perturbative renormalizability and classical scale invariance. The latter
requirement is not essential it simply limits the number of allowed couplings to those which are
dimensionless.

At first one might think that gauge theory is the only option for having asymptotic freedom with spin-1
fields but it turns out this is not the case, at least in Euclidean signature. The explicit computation of the
1-loop $\beta$-functions in the space of 7 couplings (corresponding to the 7 allowed operators in the
most general Lagrangian) shows that for any $N$ a finite number of asymptotically free RG flows exist,
more precisely 4 of these for $N>5$. These RG flows correspond to non-trivial perturbative and
asymptotically free quantum field theories which are not gauge theories. Straightforward large-$N$
scaling works as expected, and the qualitative features of the $N\to\infty$ model is the same as with
finite $N > 5$.

Similar questions as the one addressed in this work were discussed in the abelian case in 
\cite{Iliopoulos:1980zd} and rather qualitatively for the non-abelian case in \cite{Forster:1980dg}.

The most general Lagrangian for the study of spin-1 fields is given in section \ref{lagrangian}. The
$\beta$-functions are computed in section \ref{betafunctionsandrgflows} to 1-loop and the resulting RG
flows are studied as well. Asymptotically free RG flows are identified and the large-$N$ limit
is also spelled out. Finally, section \ref{conclusionandoutlook} contains our conclusions and outlook to
possible refinements and further research.

\section{Lagrangian}
\label{lagrangian}

The spin-1 fields will be labelled by $A_\mu^a$ in the adjoint representation of $SU(N)$. We seek
the most general 4-dimensional, globally $SU(N)$ and Euclidean invariant Lagrangian with at most 
two derivatives, dimensionless couplings and perturbatively
renormalizable interactions. It is straightforward to show that up to total derivatives a possible
parametrization in terms of 7 couplings, $(z,g_1,g_2,g_3,g_4,h_1,h_2)$ is,
\bea
\label{lagrangiandef}
{\mathscr L} &=& \frac{1}{2} \partial_\mu A_\nu^a \partial_\mu A_\nu^a - 
\frac{1}{2}\left(1-\frac{1}{z}\right) (\partial_\mu A_\mu^a )^2 + 
h_1 {\tilde{\mathscr O}}_1 +  h_2 {\tilde{\mathscr O}}_2 + {\mathscr V}  \nn \\
{\tilde {\mathscr O}}_1 &=& A_\mu^a A_\nu^b \partial_\mu A_\nu^c d_{abc} \nn \\
{\tilde {\mathscr O}}_2 &=& A_\mu^a A_\nu^b \partial_\mu A_\nu^c f_{abc} \nn \\
{\mathscr V} &=& \sum_{i=1}^4 g_i {\mathscr O}_i \\
{\mathscr O}_1 &=& \frac{1}{8} A_\mu^a A_\mu^b A_\nu^c A_\nu^g d_{abe} d_{cge} \geq 0 \nn \\
{\mathscr O}_2 &=& \frac{1}{8N}( A_\mu^a A_\mu^a )^2 \geq 0 \nn \\
{\mathscr O}_3 &=& \frac{1}{8N} A_\mu^a A_\mu^b A_\nu^a A_\nu^b \geq 0 \nn \\
{\mathscr O}_4 &=& \frac{1}{4} A_\mu^a A_\mu^b A_\nu^c A_\nu^g f_{ace} f_{bge} \geq 0\;, \nn
\eea
where $f_{abc}$ is the totally anti-symmetric and $d_{abc}$ is the totally symmetric tensor of $SU(N)$.
For a well-defined path integral representation $z\geq 0$ is required as well as a non-negative
potential  ${\mathscr V}$.
The requirement on $(g_1,g_2,g_3,g_4)$ for the latter to hold is non-trivial, one of the
following two conditions is necessary, 
\bea
&g_1& \geq 0\;, \qquad g_2 + g_3 \geq - g_1 ( N - 2 ) \nn \\
&g_1& \leq 0\;, \qquad g_2 + g_3 \geq - g_1 \frac{2(N-2)^2}{N-1}\;,
\eea
and any one of the following is sufficient,
\bea
\begin{array}{cccc}
g_1 \geq 0\;, &\;\; 4 g_2 + g_3 \geq 0\;,     &\;\; g_3 \geq 0\;, &\;\; g_4 \geq 0 \\
g_1 \geq 0\;, &\;\; 4 g_2 + g_3 \geq 8 g_1\;, &\;\; g_3 \geq 0\;, &\;\; 3 g_4 \geq - 2 g_1 \\
g_1 \geq 0\;, &\;\; g_2 + g_3 \geq 0\;,       &\;\; g_3 \leq 0\;, &\;\; g_4 \geq 0 \\
g_1 \leq 0\;, &\;\; g_2 + 2(N-1)g_1 \geq 0\;, &\;\; g_3 \geq 0\;, &\;\; g_4 \geq 0\;. \\ 
\end{array}
\eea
A complete set of minimal necessary {\em and} sufficient conditions is presently not known.

\section{$\beta$-functions and RG flows}
\label{betafunctionsandrgflows}

\begin{figure}
\begin{center}
\includegraphics[width=4cm]{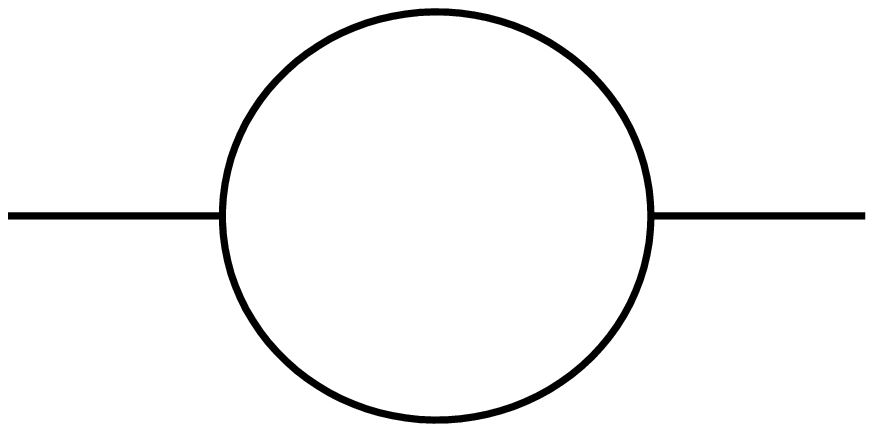} \\
\includegraphics[width=4cm]{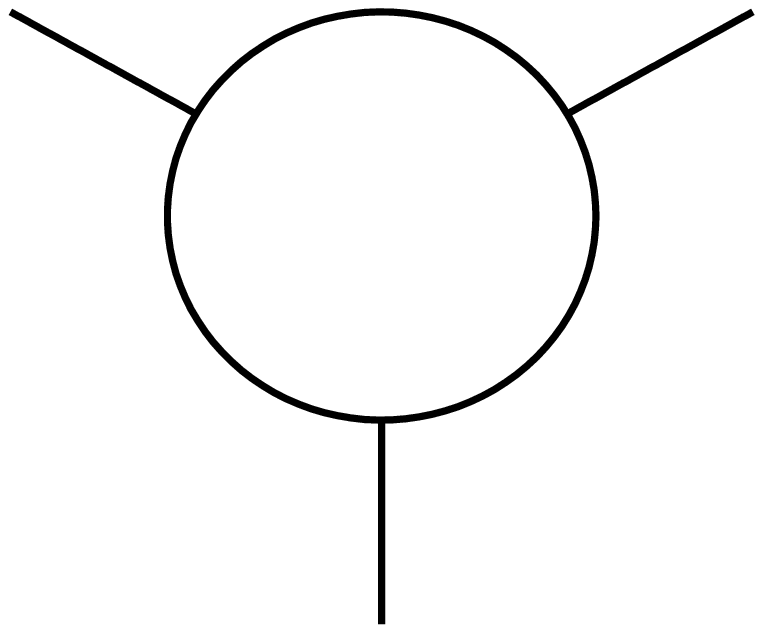} \includegraphics[width=4cm]{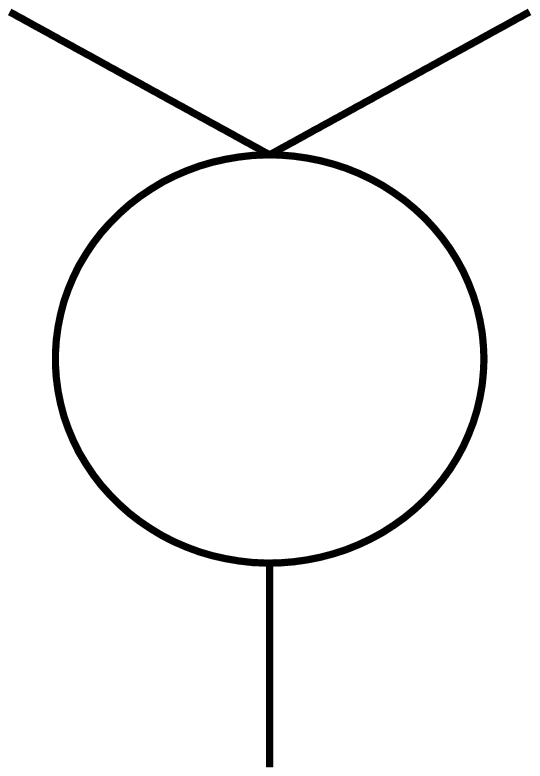} \\
\vspace{0.5cm}
\includegraphics[width=4cm]{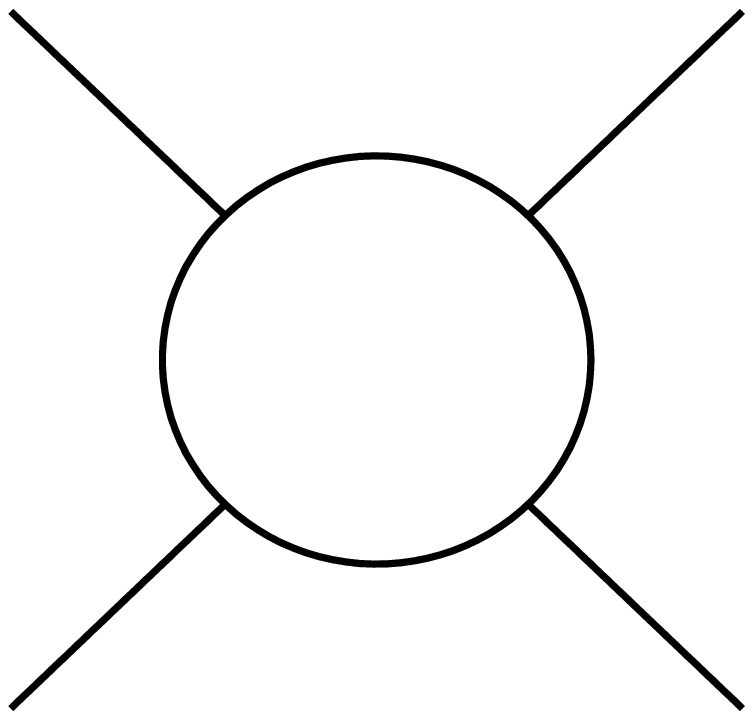} \includegraphics[width=4cm]{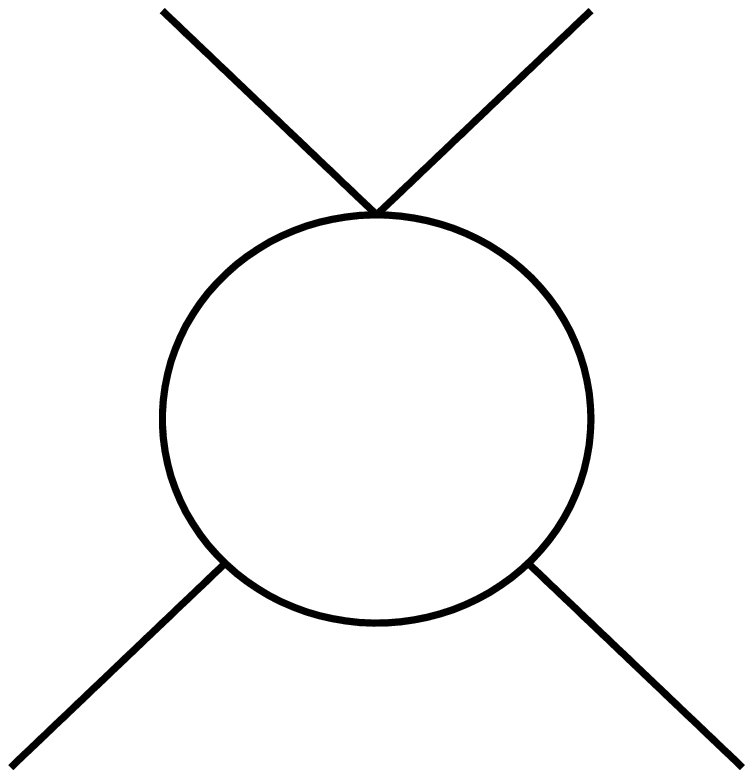} \includegraphics[width=4cm]{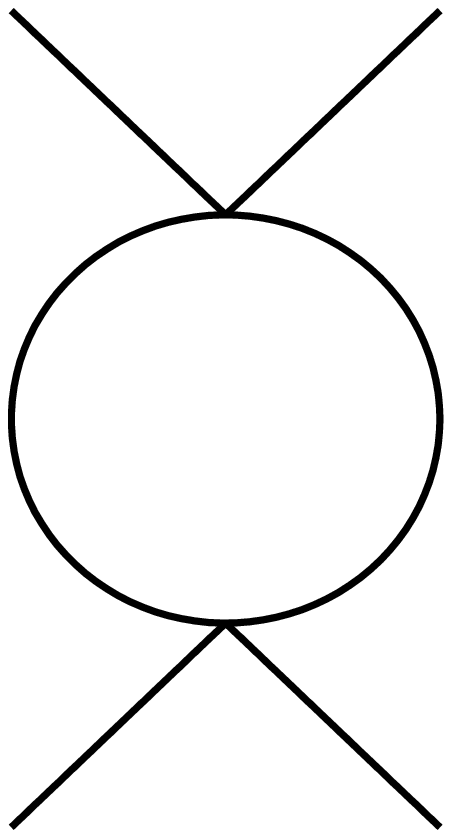} 
\end{center}
\caption{Diagrams contributing at 1-loop order in dimensional regularization. 
Rows from top to bottom: propagator, renormalization of $z$;
3-vertex, renormalization of $(h_1,h_2)$; 4-vertex, renormalization of $(g_1,g_2,g_3,g_4)$.
}
\label{diagrams}
\end{figure}

Since all possible terms allowed by symmetry are included in (\ref{lagrangiandef}), all terms are
perturbatively renormalizable and a well-defined path integral can be defined in Euclidean
signature, the $\beta$-functions of the 7 couplings can be computed in a
straightforward manner. The diagrams contributing in $\msbar$ at 1-loop are listed in figure
\ref{diagrams}. For simplicity let us introduce $g_5 = h_1^2$ and $g_6 = h_2^2$. 
Schematically, the 1-loop $\beta$-functions are,
\bea
\mu \frac{dz}{d\mu} &=& \beta_z \; = \; z L_z(g_5,g_6)\nn \\ 
\mu \frac{dg_{i}}{d\mu} &=& \beta_i \; = \; Q_{i}(g_1,g_2,g_3,g_4,g_5,g_6) \qquad i=1,2,3,4\\
\mu \frac{dg_{i}}{d\mu} &=& \beta_i \; = \; g_{i} L_{i}(g_1,g_2,g_3,g_4,g_5,g_6) \;\quad i=5,6 \;, \nn 
\eea
where $Q_{1,2,3,4}$ are quadratic monomials in the couplings with coefficients which are themselves polynomial in $z$
and $L_{z,5,6}$ are linear in the couplings and also polynomial in $z$. All expressions depend on $N$ as
well. Clearly, $z, g_5, g_6$ renormalize multiplicatively. 
The precise form of the $\beta$-functions can be found in \cite{Nogradi:2021nqm}, which were computed
with the extensive help of FORM \cite{Vermaseren:2000nd, Kuipers:2012rf, Ruijl:2017dtg}.

There is a line of Gaussian fixed points in the space of couplings given by an 
arbitrary $z$ and $g_i = 0$ for all $i=1,\ldots,6$. Clearly, $\beta_z = \beta_i = 0$ everywhere on this
line. We will be looking for RG flows which in the UV end
up on this line asymptotically. Such an RG trajectory will define a non-trivial
perturbative quantum field theory. Both $g_5$ and $g_6$ can not be identically zero, but for the sake of
simplicity let's assume $g_5 = 0$. The situation with $g_5 \neq 0$ is spelled out in detail in
\cite{Nogradi:2021nqm}. Now we are dealing with 6 couplings $(z,g_1,g_2,g_3,g_4,g_6)$ and look for RG
flows which for $\mu\to\infty$ behave as,
\bea
g_i(\mu) &\sim& 16\pi^2 \frac{C_i}{\log\frac{\mu}{\Lambda}}, \qquad i = 1,\ldots,4,6 \nn \\
z(\mu) &\sim& const\;,
\label{rgflow}
\eea
with some scale $\Lambda$ and constants $C_i$ which are subject to the 
non-trivial positivity constraint mentioned in section
\ref{lagrangian}. Assuming an asymptotically free RG flow as in (\ref{rgflow}),
clearly the ratios $r_i = g_i / g_6$ for $i=1,\ldots,4$ are constant towards the UV, $r_i \to C_i / C_6$. 
Hence our goal is to
find UV fixed points in the space $(z,r_1,r_2,r_3,r_4)$ and asymptotically free $g_6$, 
which is a straightforward exercise once the $\beta$-functions
are known explicitly. The results will be solutions of complicated polynomial equations for every $N$ and
are given in table \ref{roots}.

\begin{table}
\small
\begin{center}
\begin{tabular}{|c|c|c|c|c|c|c|c|c|}
\hline
$N$   & $z$   & $r_1$        & $r_2$        & $r_3$        & $r_4$        & $N C_6$      & ${\mathscr V}$ \\
\hline
\hline
    3 &     0 &     0.054652 &     0.122003 &     0.485317 &     0.970537 &     0.138656 &     stable \\ 
    3 &     0 &     0.064145 &     0.133021 &     0.665179 &     0.964086 &     0.137153 &     stable \\ 
    3 &     0 &    -0.647582 &    -0.580231 &     1.889786 &     1.204615 &     0.138656 &   unstable \\ 
    3 &     0 &    -0.562664 &    -0.493787 &     1.918797 &     1.173022 &     0.137153 &   unstable \\ 
    3 & 25/3  &     0.000334 &     0.079592 &    -0.251950 &     1.020083 &     0.148484 &   unstable \\ 
    3 & 25/3  &     0.010673 &     0.074642 &    -0.144563 &     1.004360 &     0.145542 &   unstable \\ 
    3 & 25/3  &    -0.108161 &    -0.028903 &    -0.034960 &     1.056248 &     0.148484 &   unstable \\ 
    3 & 25/3  &    -0.080316 &    -0.016348 &     0.037417 &     1.034690 &     0.145542 &   unstable \\ 
\hline
    4 &     0 &     0.044841 &     0.106784 &     0.351786 &     0.979028 &     0.140948 &     stable \\ 
    4 &     0 &     0.074162 &     0.083060 &     1.368389 &     0.960858 &     0.136196 &     stable \\ 
    4 & 25/3  &     0.004413 &     0.111209 &    -0.323177 &     1.013219 &     0.146900 &   unstable \\ 
    4 & 25/3  &     0.016297 &     0.243636 &    -0.344606 &     0.995511 &     0.145494 &   unstable \\ 
    4 & 25/3  &     0.017435 &     0.119096 &    -0.223217 &     0.997309 &     0.144605 &   unstable \\ 
    4 & 25/3  &     0.017931 &     0.235838 &    -0.327356 &     0.993784 &     0.145177 &   unstable \\ 
\hline
    5 &     0 &     0.042754 &     0.103223 &     0.327436 &     0.981138 &     0.141567 &     stable \\ 
    5 &     0 &     0.054311 &     1.073479 &     0.536511 &     0.957994 &     0.142046 &     stable \\ 
    5 &     0 &     0.067257 &    -0.066910 &     1.896637 &     0.967324 &     0.136857 &     stable \\ 
    5 &     0 &     0.069027 &     0.516675 &     1.600829 &     0.956705 &     0.138188 &     stable \\ 
    5 & 25/3  &     0.012566 &     0.149475 &    -0.375377 &     1.003344 &     0.145326 &   unstable \\ 
    5 & 25/3  &     0.021321 &     0.180212 &    -0.347564 &     0.993910 &     0.144298 &   unstable \\ 
\hline
\bf 6 & \bf 0 & \bf 0.041817 & \bf 0.101590 & \bf 0.316866 & \bf 0.982127 & \bf 0.141864 & \bf stable \\ 
    6 &     0 &     0.048648 &     1.137578 &     0.428569 &     0.966346 &     0.142530 &     stable \\ 
    6 &     0 &     0.059916 &    -0.214070 &     2.277709 &     0.972682 &     0.137748 &     stable \\ 
    6 &     0 &     0.062649 &     0.434621 &     2.043391 &     0.963808 &     0.138624 &     stable \\ 
\hline
\bf 7  & \bf 0 & \bf 0.041301 & \bf 0.100682 & \bf 0.311136 & \bf 0.982683 & \bf 0.142032 & \bf stable \\ 
    7  &     0 &     0.045944 &     1.161333 &     0.383774 &     0.971232 &     0.142626 &     stable \\ 
    7  &     0 &     0.054742 &    -0.321825 &     2.541816 &     0.976034 &     0.138570 &     stable \\ 
    7  &     0 &     0.057376 &     0.412019 &     2.341096 &     0.968497 &     0.139238 &     stable \\ 
\hline
\bf 10 & \bf 0 & \bf 0.040625 & \bf 0.099483 & \bf 0.303720 & \bf 0.983425 & \bf 0.142259 & \bf stable \\ 
    10 &     0 &     0.042691 &     1.184351 &     0.334451 &     0.977917 &     0.142606 &     stable \\ 
    10 &     0 &     0.047136 &    -0.495636 &     2.966144 &     0.980468 &     0.140207 &     stable \\ 
    10 &     0 &     0.048800 &     0.401667 &     2.839408 &     0.975942 &     0.140564 &     stable \\ 
\hline
\bf 50 & \bf 0 & \bf 0.040047 & \bf 0.098451 & \bf 0.297474 & \bf 0.984071 & \bf 0.142458 & \bf stable \\ 
    50 &     0 &     0.040124 &     1.198242 &     0.298567 &     0.983855 &     0.142474 &     stable \\ 
    50 &     0 &     0.040300 &    -0.680516 &     3.425269 &     0.983967 &     0.142360 &     stable \\ 
    50 &     0 &     0.040376 &     0.410710 &     3.418741 &     0.983752 &     0.142375 &     stable \\ 
\hline
\bf 100 & \bf 0 & \bf 0.040030 & \bf 0.098420 & \bf 0.297287 & \bf 0.984091 & \bf 0.142464 & \bf stable \\ 
    100 &     0 &     0.040049 &     1.198589 &     0.297559 &     0.984037 &     0.142468 &     stable \\ 
    100 &     0 &     0.040093 &    -0.686738 &     3.440904 &     0.984065 &     0.142439 &     stable \\ 
    100 &     0 &     0.040112 &     0.411281 &     3.439259 &     0.984011 &     0.142443 &     stable \\ 
\hline
$\mathbf\infty$ & \bf 0 & \bf 0.040024 & \bf 0.098409 & \bf 0.297224 & \bf 0.984097 & \bf 0.142466 & \bf stable \\
       $\infty$ &     0 &     0.040024 &     1.198704 &     0.297224 &     0.984097 &     0.142466 &     stable \\
       $\infty$ &     0 &     0.040024 &    -0.688818 &     3.446135 &     0.984097 &     0.142466 &     stable \\
       $\infty$ &     0 &     0.040024 &     0.411476 &     3.446135 &     0.984097 &     0.142466 &     stable \\
\hline
\end{tabular}
\end{center}
\caption{Non-trivial fixed points with $g_5 = 0$ for the ratios $r_i = g_i / g_6 = C_i / C_6$,
and the coefficient $C_6$; see (\ref{rgflow}). The
last column indicate whether the potential ${\mathscr V}$ is stable or not.
For $N > 5$ there is a unique fixed point for which ${\mathscr V} \geq 0$ and is stable in the RG-sense
in the $z=0$ plane, these are shown in bold. 
}
\label{roots}
\end{table}

It is clear from table \ref{roots} that for any $N$ there is a finite number of asymptotically free RG
flows. Once an RG flow is identified it may be characterized by the stability or instability of
${\mathscr V}$ and also by its stability in the RG sense.

The point $z=0$ is always a fixed point and the only fixed point for $N>5$. The $N\leq 5$ cases are
qualitatively different from $N>5$ also in the sense that in the latter case there is a unique
fixed point in the $z=0$ plane which is stable in the RG sense. Furthermore, all fixed points for $N>5$
correspond to a stable potential ${\mathscr V}$. Fixed points which correspond to a stable ${\mathscr V}$
lead to perfectly well-defined perturbative quantum field theories of spin-1 fields, which are not gauge
theories. Those which are stable in the RG sense as well are insensitive to small deformations, as usual.
The $z=0$ fixed points can be interpreted as having a constraint $\partial_\mu A_\mu^a = 0$ because of
the appearance of the coupling $1/z$ in (\ref{lagrangiandef}). As a result the original 4 degrees of
freedom are reduced to 3. Note that the $\partial_\mu A_\mu^a = 0$ constraint has nothing to do with
gauge fixing since gauge invariance is not present to begin with. The constraint arose dynamically from
the nature of the particular UV fixed points. 

It should be noted that we have been working in Euclidean signature and Wick rotation back to a unitary 
theory in Minkowski
space time is not possible. This is because, as is well-known, gauge invariance is required to kill off
the negative norm states which is of course not present on any of the RG flows considered here. In order
to study how gauge symmetry emerges in a perturbative treatment such as ours, one must include ghost
fields; for more details see \cite{Nogradi:2021nqm}.

Another aspect of table \ref{roots} is the smoothness of the large-$N$ limit. Similarly to the situation
in gauge theory the $N\to\infty$ limit is performed at constant $N g_i$. The fixed point ratios $r_i =
C_i / C_6$ have well-defined large-$N$ limits of course and so does $NC_i$. Qualitatively all $N>5$ cases
are similar, the strict $N\to\infty$ limit only makes some of the ratios between different fixed points
degenerate. The 4 fixed points only differ in $r_2$ and $r_3$ in this limit and one of them is stable in
the RG sense.

\section{Conclusion and outlook}
\label{conclusionandoutlook}

In this work a seemingly simple QFT question was posed: what is the most general QFT describing a
set of spin-1 fields with global $SU(N)$ invariance. The RG phase space was mapped out in the 1-loop
approximation and a finite number of asymptotically free RG flows were found for any $N$. More precisely,
only classically scale invariant couplings were considered, i.e. dimensionless couplings. Note that in
this case scale invariance does not imply conformal invariance 
\cite{Jackiw:2011vz,ElShowk:2011gz,Nakayama:2013is}. If dimensionful couplings are allowed, but global
$SU(N)$ invariance is still imposed, a mass term can be added to the Lagrangian,
\bea
{\mathscr L}_m = \frac{m^2}{2} A_\mu^a A_\mu^a\;.
\eea
The perturbative expansion of the corresponding anomalous dimension is beyond the scope of the present
work but would be interesting to work out in the future.

Similarly, a worthwhile extension of the present work would be the calculation of the $\beta$-functions
to 2-loops or more. Since asymptotic freedom can be established by the 1-loop calculation alone, it is
expected that the main conclusion will not change, namely that for any $N$ well-defined, asymptotically
free, perturbative Euclidean quantum field theories exist, which are not gauge theories.

\end{document}